\documentclass[aps,prx,showpacs,twocolumn,groupedaddress]{revtex4-1}
\usepackage{graphicx,graphics,color,epsfig}
\usepackage{amsmath}
\usepackage{amssymb}
\usepackage{amsfonts}
\usepackage{amsfonts}
\usepackage{epstopdf}
\usepackage{bm}
\usepackage{times,xspace}
\usepackage{color}

\begin{document}


\title{Imaging the formation of high-energy dispersion anomalies in the actinide UCoGa$_5$}

\author{Tanmoy Das, Tomasz Durakiewicz, Jian-Xin Zhu, John J. Joyce, John L. Sarrao, Matthias J. Graf}

\affiliation{Theoretical Division, Los Alamos National Laboratory, Los Alamos, NM 87545, USA}

\date{\today}

\begin{abstract}
 We use angle-resolved photoemission spectroscopy (ARPES) to image the emergence of substaintial dispersion anomalies in the electronic renormalization of the actinide compound UCoGa$_5$ which was presumed to belong to a conventional Fermi liquid family. Kinks or abrupt breaks in the slope of the quasiparticle dispersion are detected both at low ($\sim$130~meV) and high ($\sim$1~eV) binding energies below the Fermi energy, ruling out any significant contribution of phonons. We perform numerical calculations to demonstrate that the anomalies are adequately described by coupling between itinerant fermions and spin fluctuations arising from the particle-hole continuum of the spin-orbit split $5f$ states of uranium. These anomalies are
resemble the `waterfall' phenomenon of the high-temperature copper-oxide superconductors, suggesting that spin fluctuations are a generic route toward multiform electronic phases in correlated materials as different as high-temperature superconductors and actinides.
\end{abstract}


\keywords{Subject areas: angle-resolved photoemision spectroscopy, heavyy-fermion, correlation strength}

\maketitle

\section{Introduction}

The coupling between electrons and elementary excitations, originating from lattice or electronic degrees of freedom, can
drive the formation of new emergent phases, such as magnetism and superconductivity. For rare-earth and actinide $f$-electron systems, the development of low-energy fermionic excitations with heavy electron mass is complicated by the interactions between $f$ electron spins and those of itinerant electrons. They often exhibit an interplay between Kondo physics, magnetism, and superconductivity \cite{Hewson}. So far it has been thought that this physics is not at play in UCoGa$_5$ with a tetragonal crystal structure and metallic ground state~\cite{Grin1986}. In fact, it has been dubbed a {\em `vegetable'}, because of its lack of heavy mass \cite{Ikeda2003,Onuki2008},  other competing orders \cite{Grin1986,Noguchi1992,Troc2004,Moreno2005}, and showing of a spin-lattice relaxation rate that obeys the Korringa law of a conventional Fermi liquid at low temperatures \cite{Kambe2007}. However, about $\sim 75$ K it violates the Korringa law by exhibiting a quadratic temperature dependence without any sign of magnetic ordering.
The simple lack of ordering is often contrasted with the rich properties of the isostructural Pu{\it M}Ga$_5$ and PuCoIn$_5$ compounds ({\it M}=Co, Rh) \cite{Dassf}. Here, we show that UCoGa$_5$ is not a typical Fermi liquid. It is rather anomalous as evidenced by the drastic renormalization of the electronic dispersion at low and high binding energies when compared with {\it ab-initio} electronic bands. These kinks resemble the `waterfall'  phenomenon of cuprates \cite{Graf2007,Xie2007,Valla2007}  and point toward a common mechanism of spin fluctuations in both $d$ and $f$ electron systems.

The actinide systems remain poised between the strong and weak coupling limit of Coulomb interaction. Therefore they offer significant tunability across several correlated states of matter. The effective Coulomb repulsion $U$ of the $5f$ electrons is not strong enough to localize all of them, yet, it is sufficient to slow them down so they acquire a moderately increased mass near the Fermi level. The central puzzle is how does this common low-lying heavy electronic state transform into stable ground states which differ dramatically within the family~\cite{Moreno2005,UPtGa5}. An understanding of these differences hinges on knowing the nature of the exchanged bosons that dress the bare electrons to become heavy quasiparticles and ultimately drive the system into a particular stable ground state. The choice of UCoGa$_5$ for this study allows us to rule out any possible intervention of competing orders. Thus it offers a clean approach to unravel the nature of the exchanged boson in the actinide family and related heavy-fermion systems.

\section{Experiment}

\begin{figure*}[top]
\centering
\rotatebox{0}{\scalebox{.6}{\includegraphics{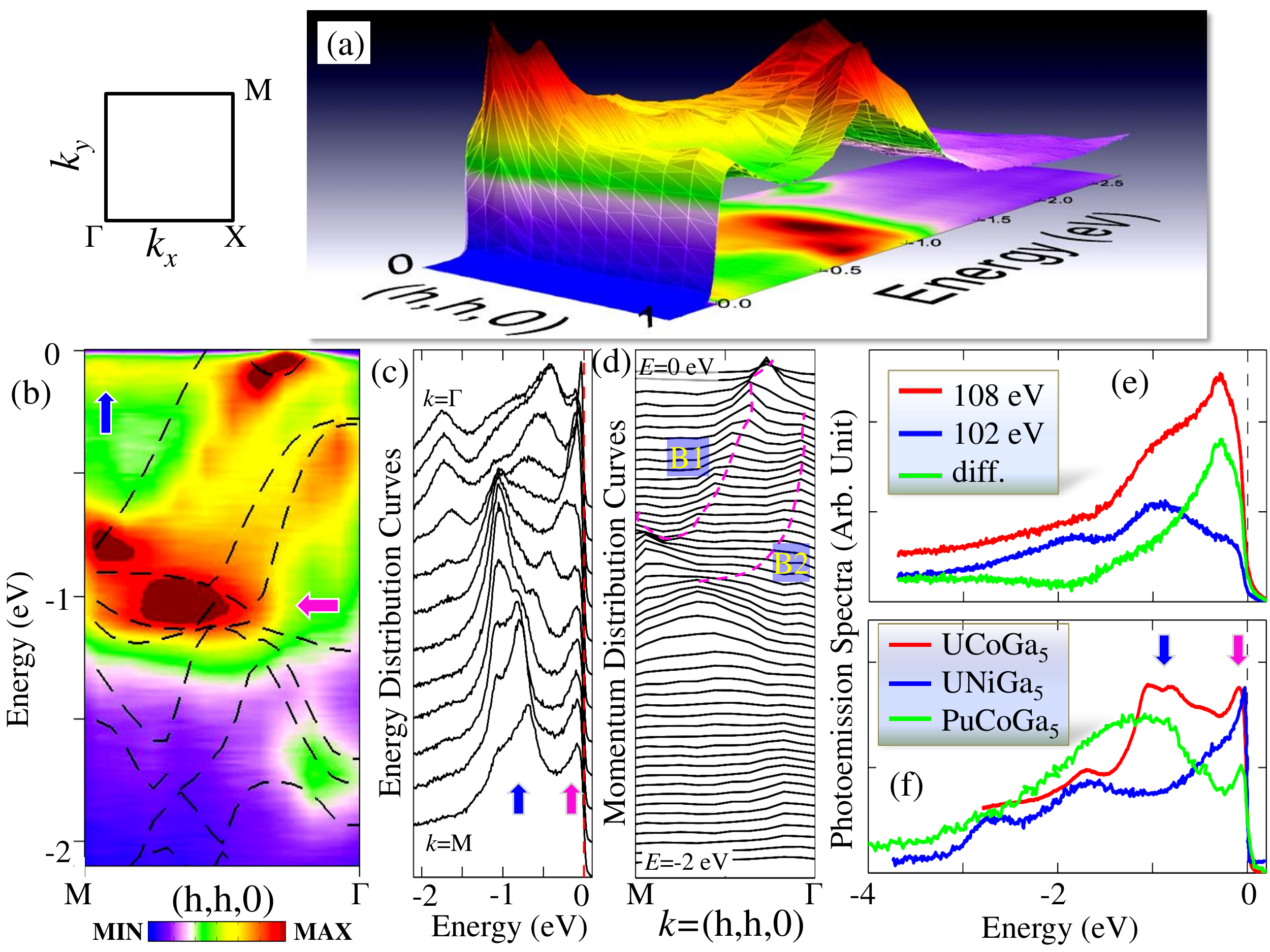}}}
\caption{{Dispersion and measured ARPES spectral function anomaly.}
({a}) 3D intensity map of UCoGa$_5$ along the M$\rightarrow\Gamma$ direction $(h,h,0)$ in the Brillouin zone. ({b}) Same data plotted as 2D contour map and compared to  the corresponding {\it ab-initio} GGA electronic band-structure dispersions (black dashed lines). Arrows indictate two quasi-non-dispersive energy scales of higher intensities. All the ARPES data are collected at the photon energy of 46~eV, with energy resolution better than 20~meV. ({c}) The EDCs are plotted for fixed momenta. The  curves from bottom to top are chosen for equally spaced momentum points from M to $\Gamma$. Arrows have same meaning as in ({a}). ({d}) MDCs for fixed binding energy. Bottom to top curves are chosen from $E$=-2.1~eV to $E$=0~eV. The dashed lines are guide to the eyes for two extracted low-lying   dispersions of anomalous character, analyzed in Fig.~2. ({e}) Photoemission spectrum (PES) of UCoGa$_5$ at photon energies $h\nu$=108~eV (red) and  $h\nu$=102~eV (blue). Their difference curve (green) reveals the dominant $5f$ character~\cite{Yeh1985}.  ({f}) Partial PES of UCoGa$_5$ (momentum integrated between $\Gamma$ and M) compared with two related actinides. UNiGa$_5$ and PuCoGa$_5$ exhibit antiferromagnetic \cite{UPtGa5} and superconducting \cite{Curro} ground states, respectively. All data are normalized arbitrarily for ease of comparison.}
\label{arpes}
\end{figure*}

We use ARPES to image the momentum and energy dependence of the electronic dispersions and thereby reveal the renormalization of itinerant bands. Details of the ARPES experiments and calculations are described in the supplementary material ({SM}\cite{SM}). Figure~1 reports our main experimental result of UCoGa$_5$. A quick visual inspection of the spectral function in Fig.~\ref{arpes}(a) and \ref{arpes}(b), (3D and 2D rendering of the same data, respectively), shows extensive spectral weight redistribution both in momentum and energy space. The dispersion (traces of the intensity peaks), in Fig.~\ref{arpes}(b), reveals a drastic departure of the quasiparticle states from the {\it ab-initio} electronic structure calculations
(black and dashed). More importantly, the associated quasiparticle width at the peak positions is significantly momentum and energy dependent. Unlike the renormalized quasiparticle dispersion relation, which is a genuine manifestation of many-body correlation effects, the anomaly in the lifetime may have many intrinsic and extrinsic origins and is more difficult to assign to a particular mechanism~\cite{Yeh1985} [see SM\cite{SM} on the extrinsic background]. We notice that while bare bands, computed within generalized gradient approximation (GGA) based density functional theory, are present in the entire energy and momentum region, the spectral intensity is accumulated mainly at two energy scales. Such a connection between low and high energy scales is analogous to the so-called `waterfall' or high-energy-kink feature observed in single-band cuprates~\cite{Graf2007,Xie2007,Valla2007}, where the band bottom lies at $\Gamma$, while here it is shifted to the M point.

The anomaly is markedly different in the energy and momentum space, which is a hallmark feature of correlated electron states. To expose this distinction, we present several energy-distribution curves (EDCs) of the ARPES intensity at several representative fixed momenta in Fig.~\ref{arpes}(c), and momentum-distribution curves (MDCs) at several fixed energy points in Fig.~\ref{arpes}(d). The peaks in the EDCs display prominent dispersionless features at two energy scales. The lowest energy peak persists at all momenta around -80~meV and gradually becomes sharper close to the $\Gamma$ point. This low-energy feature reveals the formation of the long-lived renormalized quasiparticle. We will show below that the renormalization phenomena can be quantified to originate from the interaction of electrons with spin fluctuations, even without invoking the more traditional Kondo physics of heavy fermions. The high-energy feature around $-1$~eV  is considerably broad in both energy and momentum space. In momentum space it attains a large width, demonstrating that these states are significantly incoherent.  By comparing the experimental dispersion with its {\it ab-initio} counterpart, we can convincingly draw the conclusion that these incoherent states are created by spectral-weight depletion near $-500$~meV due to coupling to spin-fluctuations. To gain further confidence about the correlation origin of these band anomalies, we note that such a redistribution of the spectral weight does not occur uniformly for the non-interacting bands; for instance the spectral weight is strongly suppressed between $-2$~eV to $-1.5$~eV near the M point, albeit several bands are present in this part of the Brillouin zone. By comparing the MDCs and EDCs, we immediately see that the spectrum is less dispersive as a function of momentum, while it disperses strongly with energy.

Figure~\ref{arpes}(e) shows the quasiparticle peak pushed below the Fermi level by approximately 290 meV in the integrated photoemission spectrum (PES). The difference curve (green line) between measurements at 108 eV and 102 eV photon energies reveals that the quasiparticle states in the vicinity of the Fermi level are predominantly of $5f$ character~\cite{Yeh1985}. The partial PES that comes from the $\Gamma\to$M momentum direction, is obtained by integrating the spectrum in Fig.~\ref{arpes}(b) and shown by the red line in Fig.~\ref{arpes}(f). Interestingly, it displays a characteristic peak-dip-hump feature, which is generic for other isostructural actinides~\cite{Joyce2003,Troc2004} (also shown here), as well as for $\delta$-Pu~\cite{AJArko00} and copper-oxide superconductors~\cite{Graf2007,Xie2007,Valla2007} (not shown).
Unlike UCoGa$_5$, the actinides UNiGa$_5$ and PuCoGa$_5$ exhibit antiferromagnetic \cite{UPtGa5} and superconducting \cite{Curro} ground states, respectively.

\section{Analysis}

\begin{figure}[top]
\hspace{-0cm}
\centering
\rotatebox{0}{\scalebox{.35}{\includegraphics{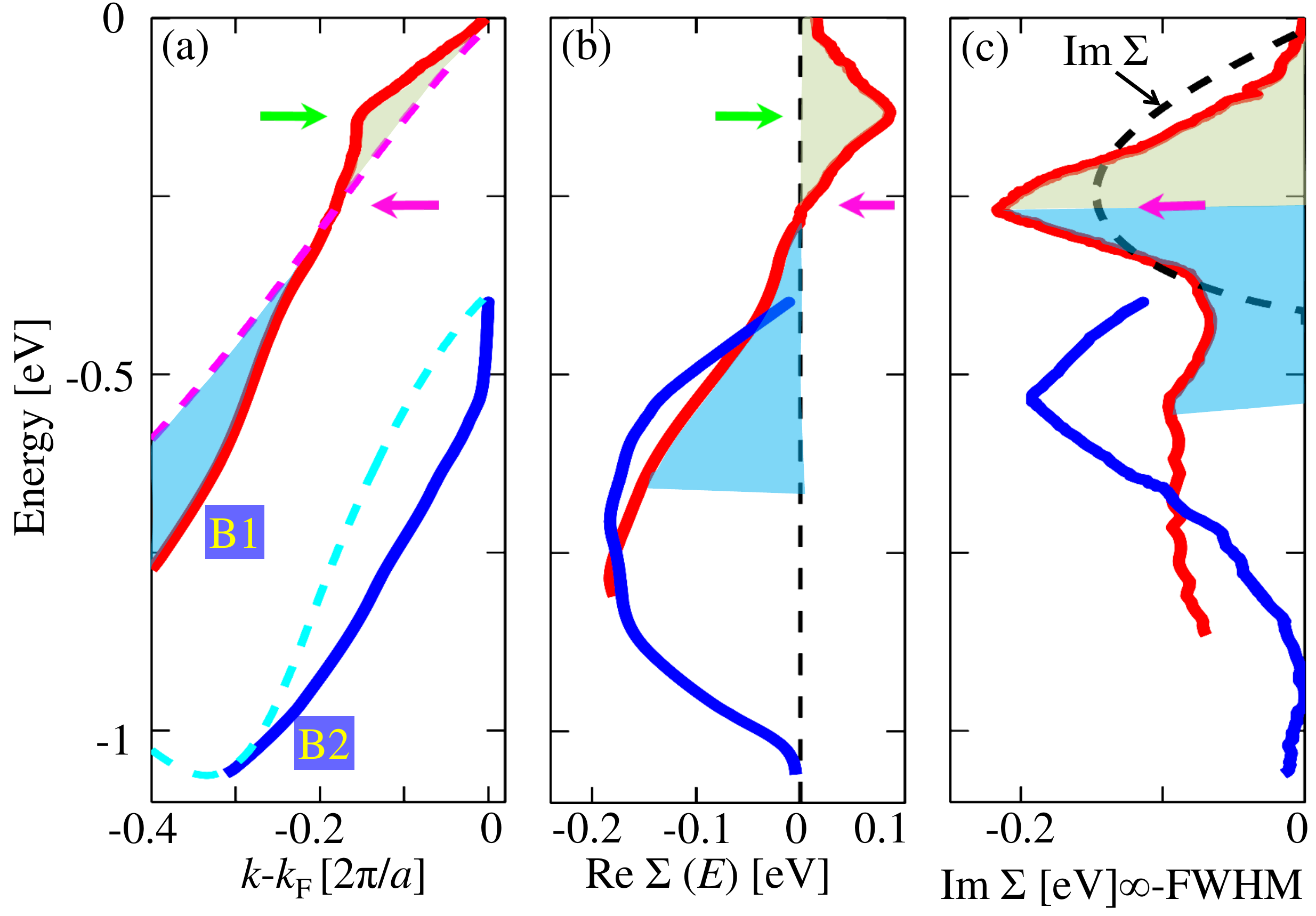}}}
\caption{{The real and imaginary parts of the experimental self-energy.} (a) Two low-lying dispersions derived from the peak positions of the MDCs [Fig.~\ref{arpes}(c)] along M$\rightarrow\Gamma$. The momentum axis is offset by $k_F$ for band B1. Dashed lines of same color give the {\it ab-initio} dispersion (offset from their corresponding Fermi momenta values) which help expose the degree of anomaly for each dispersion with respect to their linear dispersion features. (b) The real part of the self-energy measures the deviation of each dispersion curve in (a) from its corresponding  GGA value. Green arrow marks the location of  the low-energy kink. (c) The inverse of the MDCs width obtained with the help of Fig.~\ref{arpes}(c), which is proportional to the imaginary part of the self-energy, when the extrinsic background contribution is negligible. Different color fillings in all three curves (corresponding to band B1) separate the characteristic energy scale (marked by red arrows). The imaginary part attains a peak at the energy where the real part of the self-energy changes sign. To demonstrate this behavior we calculate ${\rm Im}\,\Sigma$ from ${\rm Re}\,\Sigma$ in (b) and plot is as black dashed line in (c). }
\label{se}
\end{figure}

In order to explain the correlation aspects of the dispersion anomalies, we extract two quasiparticle dispersion branches by tracing the locii of the MDC peaks as shown in Fig.~\ref{se}(a). To disentangle subtle features in the dispersion, the common practice is to observe the departure of the dispersion with respect to a featureless one, $\xi_{n{\bm k}}$ (shown by dashed lines of same color). A visual comparison reveals a sharp kink in the metallic actinide UCoGa$_5$, that is, a change in the slope of the dispersion. The low-energy kink lies around $-130$ meV (indicated by green arrow), which is higher than that in cuprates ($-70$ meV) \cite{LanzaraLEK}. On the other hand, the kink observed in the actinide USb$_2$ is at much lower energy, and thus is most likely caused by electron-phonon coupling~\cite{Durakiewcz2008}. Furthermore, the high-energy kink of the `waterfall' ($-1$ eV) poses a ubiquitous anomaly as in cuprates \cite{Graf2007,Xie2007,Valla2007}. On the basis of these comparisons, we deduce that the present kink lies well above the phonon energy scale of $\sim$30 meV for metallic UCoGa$_5$~\cite{Piekarz2005}, and hitherto provides a novel platform to study the evolution of spin fluctuations in $f$-electron systems. It it known that the role of phonon vanishes in the high-energy scales of present interest in conventional metals, unlike, for example, in quasi-one-dimensional blue and red bronzes where Peierls instability can shift the phonon energy scale by 10-15 times.\cite{Mitrovic2004, Grioni2009}

In standard quantum field theory notation, the coupling between bosonic excitations and fermionic quasiparticles is defined by a complex fermionic self-energy $\Sigma_{n}({\bm k},\omega)$ function. The real and imaginary parts of the self-energy  cause dispersion renormalization and quasiparticle lifetime broadening, respectively. This information can be derived from the ARPES spectra by tracing the peak positions of the $n$-th band ${\bm k}_{n}$ as
\begin{eqnarray}
{\rm Re}\, \Sigma_n({\bm k}, \omega) &=& \omega-\xi_{{\bm k},n}
\\
{\rm Im}~\Sigma_n ({\bm k}, \omega)&\propto& 1/{\rm FWHM}_{MDC}.
\end{eqnarray}
Here $\xi_{{\bm k},n}$ is the GGA dispersion with respect to its corresponding Fermi energy. The obtained results for  ${\rm Re}\,\Sigma$ and the inverse full width at half maximum (FWHM) of MDC are given in Fig.~\ref{se}(b) and \ref{se}(c), respectively, for the two extracted bands shown in Fig.~\ref{se}(a). Taking advantage of our high-precision data, we expose two energy scales in ${\rm Re}\,\Sigma$ (marked by arrows) which are intrinsically linked to the quasiparticle lifetime $\tau=\hbar/(2{\rm Im}\,\Sigma)$. In addition to the peak at the kink energy, ${\rm Re}\, \Sigma$ possesses a sign reversal around 260 meV. This is an important feature which imposes the constraint that the corresponding ${\rm Im}\, \Sigma$ should yield a peak exactly at the same energy, due to the Kramers-Kronig relationship. This is indeed in agreement with our findings for UCoGa$_5$ in Fig.~\ref{se}(c), which exhibits a sharp peak in the extracted broadening, exactly where ${\rm Re}\,\Sigma$ changes sign (marked by green arrow). It is worthwhile to mention that in ARPES measurements, the extrinsic source of quasiparticle broadening is typically known to be quasi-linear in energy~\cite{Shirley1972}, while the presence of a peak is a definitive signature of the intrinsic origin of broadening.

\section{Numerical Calculations}

\begin{figure*}[top]
\hspace{-0cm}
\centering
\rotatebox{0}{\scalebox{.6}{\includegraphics{{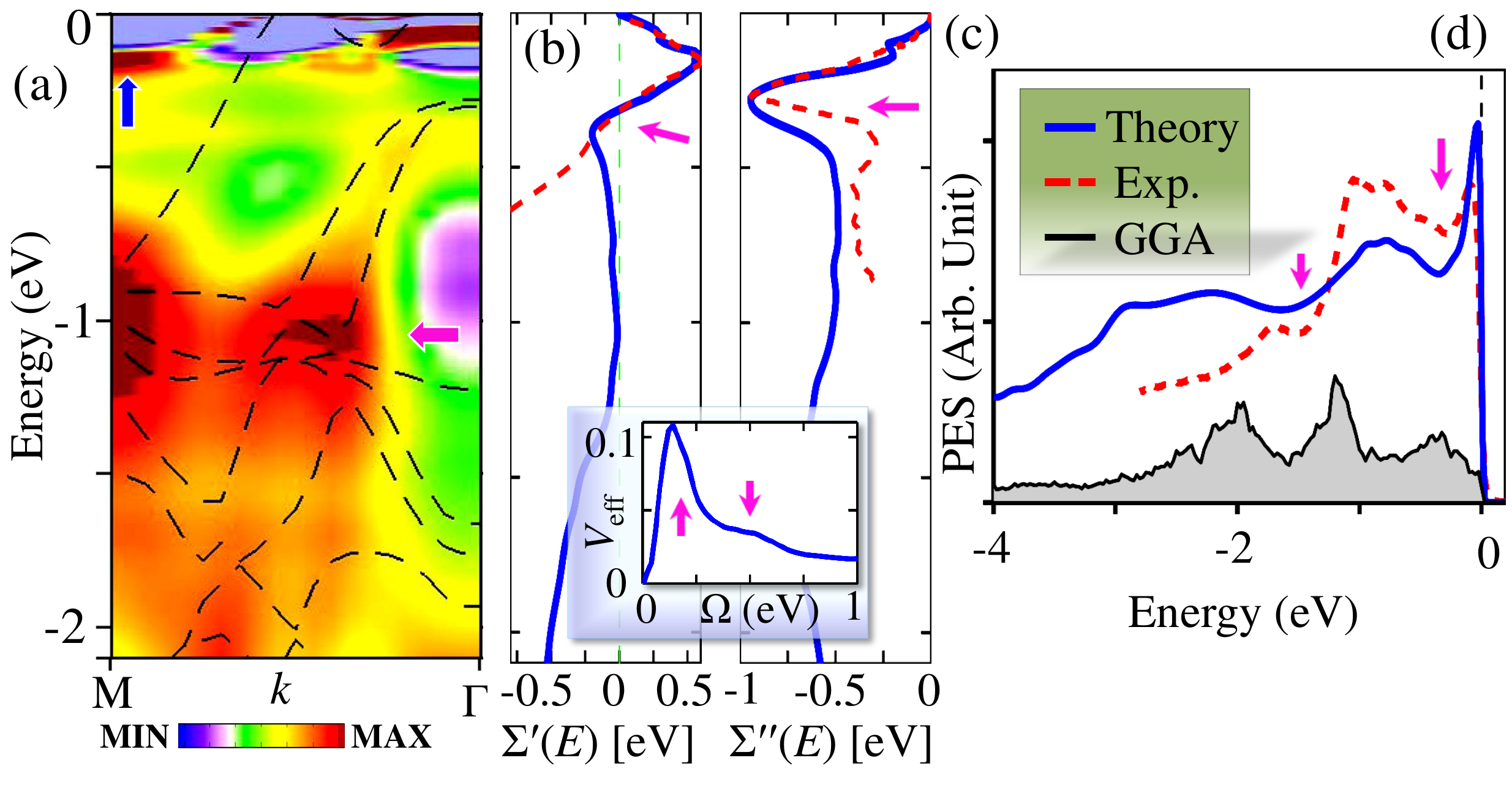}}}}
\caption{{Theoretical study of the spin-fluctuation dressed dispersion anomaly.} (a) Spectral function intensity map versus energy and momentum along M$\rightarrow\Gamma$. Black lines are the corresponding {\it ab-initio} bands.  Arrows mark the two energy scales of highest intensity, as in Fig.~\ref{arpes}(a). (b) The real part of the momentum and orbital averaged self-energy (blue line), compared with the experimental data (red-dashed line). The experimental curve is multiplied by a factor of 5 for better visualization, because it is derived from a linear slope. (c) The imaginary part of the self-energy shows a peak at the same energy where the experimental linewidth also has a peak. (d) Computed partial PES (integrated over momentum between $\Gamma$ and M) shows multiple peaks, as in the experimental curve (red dashed lines, taken from Fig.~\ref{arpes}{e}). The black line is the corresponding (normalized) GGA density of states.  The red arrows mark the depletion of states, originating from the spin fluctuation interaction.   }
\label{th}
\end{figure*}

Encouraged by the aforementioned analysis, we proceed with an {\it ab-initio} calculation to delineate the nature of the elementary excitations responsible for this anomaly. As mentioned before, both the energy scales and the strength of electron-phonon interaction \cite{Maehira2003,Opahle2004} is insufficient to capture our observed dispersion anomalies. We demonstrate that the electron-electron interaction due to dynamical spin fluctuations of exchanged bosons can describe the data. In UCoGa$_5$  strong spin-orbit coupling from relativistic effects enables substaintial spin fluctuations whose feedback  results in an increase of the electron mass and a shortening of the quasiparticle lifetime. We perform numerical calculations of the spin-fluctuation interaction potential (defined here as $V_{eff}$) and resulting self-energy $\Sigma_{n}({\bm k},\omega)$ within spin-fluctuation coupling theory. Our intermediate coupling approach constrains the value of the Coulomb potential $U$ to be of the order of the {\it ab-inito} band width $\sim 1$~eV,
for details see \cite{SM}.

The computed results are presented in Figs.~\ref{th}(b) and \ref{th}(c). A close examination of all these spectra reveals essentially two energy scales tied to the electronic structure of UCoGa$_5$. In this system, the spin-orbit coupling of U atoms splits the degeneracy in the 5$f$ states by $\sim500$~meV~\cite{Dassf}, which stipulates a strong peak in $V_{eff}$, and Im~$\Sigma$ (correspondingly Re~$\Sigma$ changes sign at this energy owing to the Kramers-Kronig relation). The second peak develops at a higher energy from the transition between unfilled $5f$ states and occupied $3d$ states of the Co atoms. Note that both energy scales are clearly discernible in $V_{eff}$ (red arrows in the insert) and shift to higher energy in ${\rm Im}\,\Sigma$ due to the effects of dispersive bands. The peaks (dips) in ${\rm Im}\,\Sigma$ correspond to electronic spectral-weight dips (peaks) in ARPES as shown in Fig.~\ref{th}(a). Importantly, the spin fluctuations  also adequately explain the quasi-dispersionless renormalized feature around -80~meV and the spectral-weight peak-dip-hump features (marked by arrows), as observed in the ARPES spectra in Fig.~\ref{arpes}(b), although some discrepancies exist. The momentum-integrated spectrum in Fig.~\ref{th}(d) further demonstrates the presence of both dips in the experimental and theoretical spectra. We note that the calculated self-energy corrections only shift the GGA Fermi surface (FS) maps in the $k_z=0$ plane. The computed FS maps are in accord with de Haas-van Alphen (dHvA) measurements, except for the tube-like FS around ($\pi/2,\pi/2,0$) \cite{QO}.
Furthermore, our GGA calculation agrees with previous electronic structure calculations \cite{Maehira2003,Opahle2004}, except for the detailed shape of the tube-like FS. By turning on the self-energy the disconnected tube-like FSs become connected along the $k_z$ direction, as shown in the SM \cite{SM}, signaling a topological transition between a closed and open FS due to a small shift in Fermi energy.

Next we turn to estimating the degree of correlations in this material. The electron-phonon and spin-fluctuation coupling strength $\lambda$ can be extracted from the slope of the self-energy in the low-energy region as  ${\rm Re}\,\Sigma_{n}({\bm k},\xi_{n}({\bm k}_F))=-\lambda\xi_{n}({\bm k}_F)$. Our theoretical estimate of the spin-fluctuation coupling constant  is $\lambda_{SF}=1.0$. Experimentally, an estimate of $\lambda$ at the Fermi level is obtained by comparing the measured Sommerfeld coefficient with respect to its non-interacting value, $\gamma \approx (1+\lambda)\gamma_{0}$. The reported measured values of $\gamma$ are between $\gamma=$10-21~mJ/mol~K$^{2}$~\cite{Moreno2005,Noguchi1992}.
Our GGA calculated value of $\gamma_{0}$=5.8~mJ/mol~K$^{2}$ agrees well with the literature
\cite{Opahle2004,Maehira2003} and thus
gives an experimental estimate of $\lambda_{exp}\approx 0.7-2.6$.  Therefore, the combination of spin fluctuations, $\lambda_{SF}=1.0$, and low-energy electron-phonon coupling,
$\lambda_{ph}\sim$0.7 for PuCoGa$_5$ \cite{Metoki2006,Piekarz2005}, adequately describes the observed electronic mass renormalization in UCoGa$_5$.

\section{Conclusions}

Our intermediate Coulomb-coupling approach offers an opportunity to address the complex nature of the  dynamic band renormalization in actinides. This is not unexpected for materials with itinerant electrons, since similar theories based on the coupling between spin fluctuations and fermionic quasiparticles are among the leading contenders for explaining the origin of high-temperature superconductivity \cite{Scalapino} and other emergent states of matter in $d$- and  $f$-electron systems \cite{Moriya,Moriya1985}.
In addition, spin fluctuations in $5f$-electron systems are a manifestation of relativistic effects due to spin-orbit split states of order $\sim$500~meV in UCoGa$_5$. The emergence of a dispersionless band around -80~meV may be related to the break-down of the Korringa law for the spin-lattice relaxation rate observed above $\sim$75 K \cite{Kambe2007}. In fact, spin-orbit coupling may provide a novel electronic tunability for the interaction strength and characteristic frequency of the mediating boson. The results for the self-energy will motivate efforts to identify generic dispersion anomalies and `waterfall' physics in a larger class of actinide and heavy-fermion compounds.

\begin{acknowledgments}
We thank A.V. Balatsky, F. Ronning and E.D. Bauer for discussions. Work at the Los Alamos National Laboratory was supported by the U.S.\ DOE under Contract No.\ DE-AC52-06NA25396 through the Office of Science (BES) and the LDRD Program. The SRC is supported by the NSF under Award No.\ DMR-0084402. We acknowledge computing allocations by LANL's Institutional Computing and NERSC under Contract No.\ DE-AC02-05CH11231.
\end{acknowledgments}

\section{Supporting Material}

\section{Measurements}
The photoemission measurements of the electronic structure of UCoGa$_5$ were performed in the Synchrotron Radiation Center (SRC) using high-quality single crystals grown by the flux method. The Plane Grating Monochromator beamline and a single-channel photoelectron energy analyzer were used. The base pressure in the measurement chamber was at 3$\times$10$^{-11}$ Torr with cryostat and samples at room temperature and at 2$\times$10$^{-11}$ with cryostat and samples at 12~K. The acceptance angle of the analyzer was set to $\pm$1 degree, and energy resolution at 46~eV and 108~eV was 20~meV and 35~meV, respectively. The EDC were recorded in $12.5$ meV increments at 11 equidistant steps between $\Gamma$ and M.
The sample was oriented using a Laue camera prior to the measurement. A detailed set of normal emission scans was collected in order to estimate the inner potential value, determine the correct photon energy at the $\Gamma$  point (46~eV) in the Brillouin zone, and search for $k_z$ dispersion. The observed $k_z$ dispersion of bands rules out the presence of spurious surface states resulting in flat bands below the Fermi energy. The difference between the 108 eV and 102 eV resonances is due to different cross-sections for $5f$ and $5d$ electrons, respectively. Hence it is characteristic of the strength of the $5f$ signal \cite{Yeh1985}. The U 5$d$-5$f$ resonance at 108~eV probes the midpoint between $\Gamma$ and Z  with $k_z\approx \pi/2$, while at 102~eV it probes near $k_z\approx 0.8\pi$. Single crystals of UCoGa$_5$ and UNiGa$_5$ were grown from a gallium flux  \cite{Moreno2005}.


\section{3D visualization of the ARPES spectra}
%
%
\begin{figure}[t]
\rotatebox[origin=c]{0}{\includegraphics[width=.95\columnwidth]{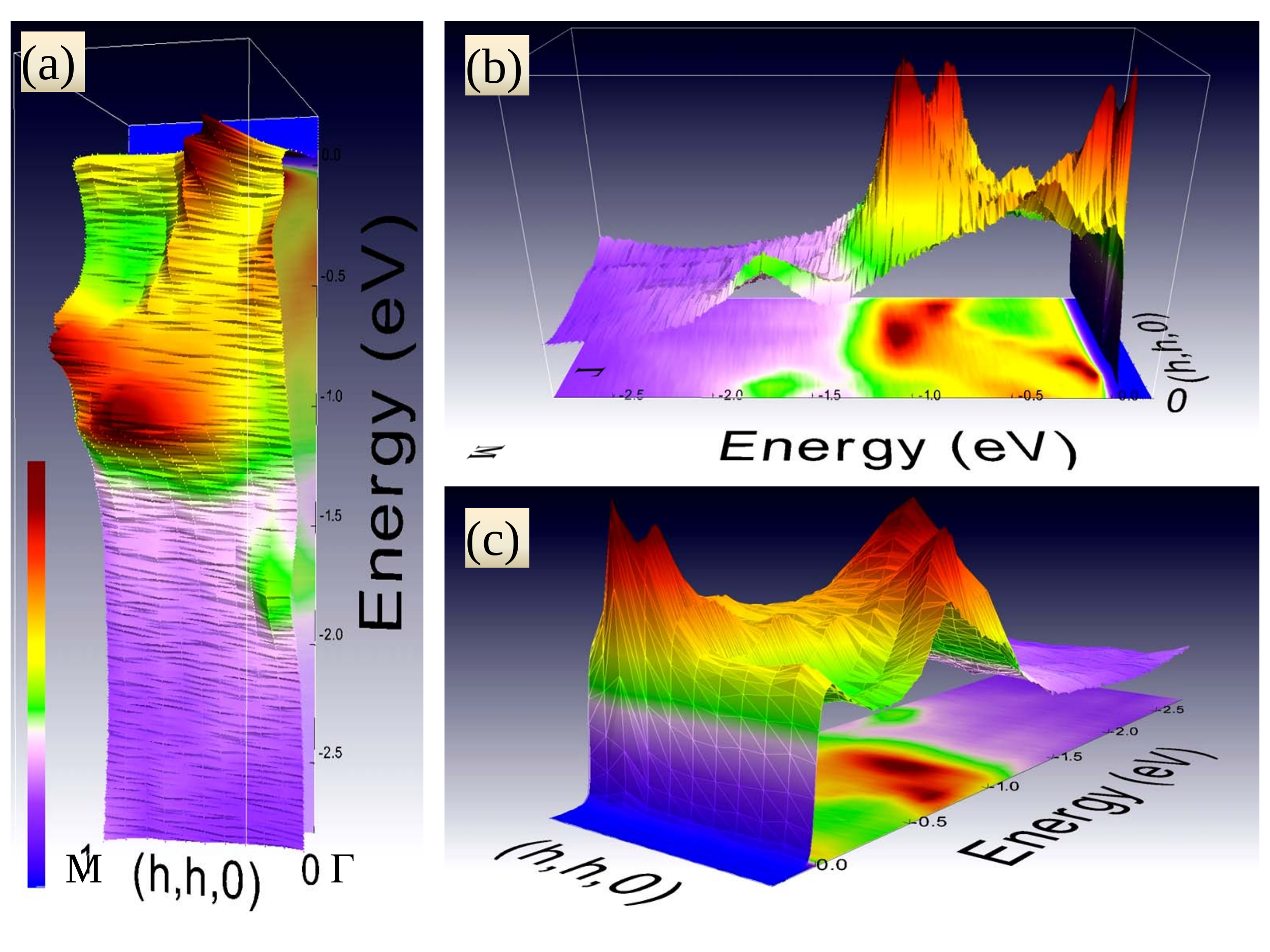}}
\caption{ {\bf Fig. S1. 3D ARPES visualization.}
 (a) Visualization of the experimental ARPES intensity contour map of UCoGa$_5$ in the Brillouin zone between  M and $\Gamma$.
 (b) Side view of the 3D surface plot of the ARPES data with projection of the contour map onto the horizontal plane.
 (c) 3D surface plot of the ARPES data with projection onto horizontal plane.
 }
\end{figure}

Figure S1 shows three different views of the experimental ARPES data in the Brillouin zone between $\Gamma = (0,0,0)$ and M$=(1,1,0)$. It facilitates a qualitative analysis of the low-energy and high-energy anomalies, kinks, waterfalls and dispersionless bands through visual inspection, while at the same time providing an overview. The intensity contour map is plotted in Fig.~S1{(a)} with small spheres indicating the measured points. The EDCs were recorded in steps of 12.5 meV at 11 momentum points between $\Gamma$ and M. Figure~S1(b) shows a 3D visualization of the data. The $z$ axis value of the surface plot is proportional to the intensity of the spectral function and clearly delineates peaks, dips and ridges. The measured data points are shown as projected points in the horizontal $xy$ plane and as vertices in the overlayed surface mesh. Figure~S1(c) shows a side view of the spectral function to emphasize the peak-dip-hump structure seen in the integrated photoemission spectrum. The intensity contour map is projected onto the horizontal $xy$ plane.

\section{Photon energy dependence of high-energy anomaly}
%
%
\begin{figure}[h]
\rotatebox[origin=c]{0}{\includegraphics[width=.95\columnwidth]{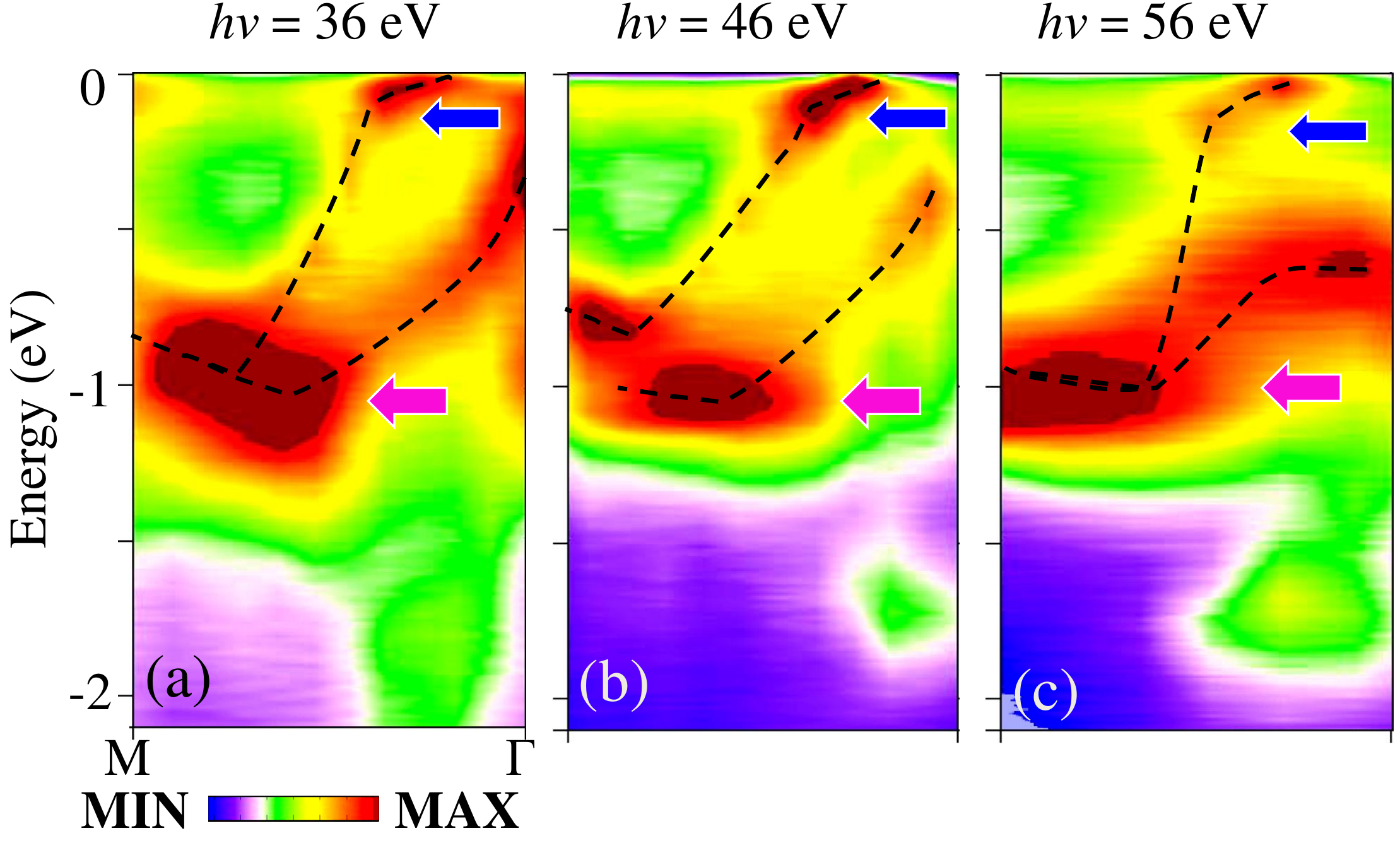}}
\caption{{\bf Fig. S2. ARPES spectra at three representative photon energies} The dashed lines are guide to the eyes about the nature of the dispersion.  }
\end{figure}

Figure~S2 demonstrates that the salient high-energy feature is a genuine manifestation of the intrinsic correlation effect, and is very much independent of the photon-energy. We chose these three representative photon energies, yet as wide as 20~eV, out of a total of 16 photon energy scans done with high resolution ARPES, in which the signature of the
high energy scale is present in all of them and also they best covers one full Brillouin zone ($\gamma$-$\Gamma$) and midpoint at $Z$.

In a separate normal emission study we have sampled over 30 photon energies between 14 eV and resonances (108eV) and the intensity variations in the energy region of interest correspond to moving through consecutive Brillouin zones, and 
not to the Co 3$p$ to 3$d$ resonances. In 5$d$-5$f$ resonance study we also clearly show that the features at 1~eV binding energy are of 5$f$ character.

All above findings clearly demonstrate that the high energy features we observe are not related to the matrix element effects, but are a direct consequence of the electronic structure renormalization. We note that the photon energy dependence of the `waterfall' physics obtained earlier for the Bi-based cuprate\cite{Inosov} is caused due to the bi-layer splitting\cite{Basak}, and thus not been seen in any other single-layered cuprates,\cite{Lanzara} or in Sr$_2$Ru2O$_4$\cite{Iwasawa} or in the present system.

\subsection{Why is the spectral function anomaly not created by ARPES matrix-element effects?}
%
%
\begin{figure}[h]
\rotatebox[origin=c]{0}{\includegraphics[width=.9\columnwidth]{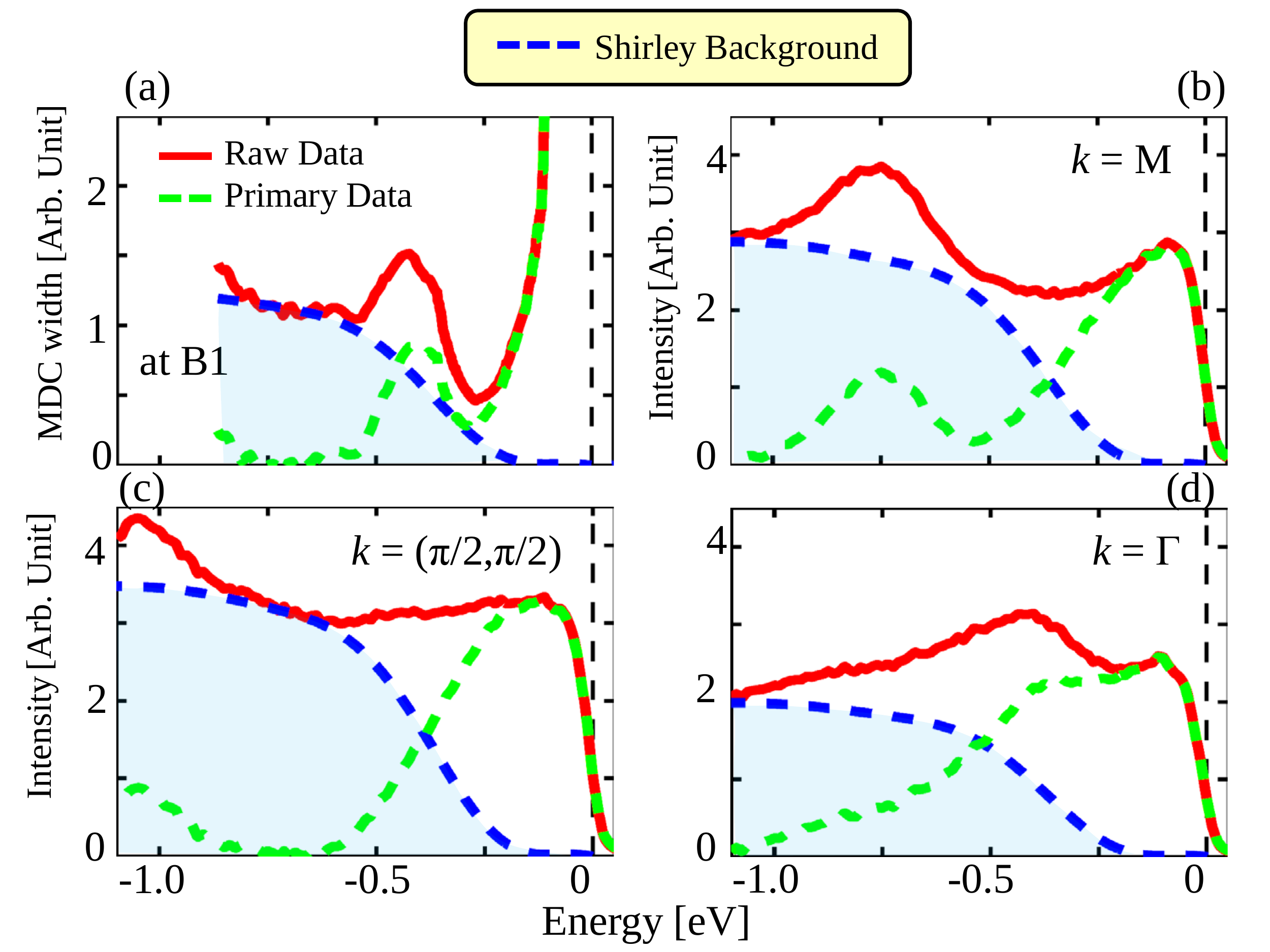}}
\caption{
{\bf Fig. S3. EDC spectra after subtracting the background matrix-element contributions.}
In all figure panels, red solid lines are the measured spectra at several representative momentum points. The blue dashed lines are the Shirley background for the corresponding spectra, while green dashed lines are the corrected spectra obtained after subtracting the background. Even after subtracting  the background, all spectra continue to exhibit the peak-dip-hump feature of our present interest.}
\end{figure}

In order to convincingly establish the fact that the spectral function anomaly observed in the ARPES spectra is a true manifestation of the many-body effects, we adopt a method originally proposed by Shirley. \cite{shirley} According to him,
the ARPES background below the energy scales of the actual valence bands becomes nearly constant. The background in the energy scales of the valence band arises entirely from valence band photoelectrons that were inelastically scattered before leaving the sample. In this method the `primary' photoelectrons (un-scattered in the photoemission process) can be separated from the background of inelastic `secondary' photoelectrons as \cite{dessau}
\begin{equation}\label{shirley}
P(E) = R(E) - B(E) = R(E) - \kappa\int_0^E P(E^{\prime})dE^{\prime},
\end{equation}
where $R(E)$ is the raw measured spectral function at constant momentum or energy distribution curve (EDC), $B(E)$ is the background, and $P(E)$ is the primary spectrum or peak of the EDC data after Shirley background correction. Here $E$ is the binding energy measured relative to the Fermi level at $E=0$. Eq.~\cite{shirley} is a self-consistent equation, which implies that the extrinsic scattering process that a photoelectron undergoes before being detected depends on its intrinsic spectral weight distribution. The proportionality constant $\kappa$ determines the fraction of primary photoelectrons that participated in the scattering process to become the secondary photoelectrons. In principle, $\kappa$ will depend on the band index. However, for the energy scale of present interest, we have demonstrated in the main text that the bands are well separated up to the energy scale of $\sim0.8$~eV along M$\rightarrow\Gamma$, and thus we assume $\kappa$ to be independent of energy, but momentum dependent.

As shown in Fig.~S3, the obtained Shirley background term is almost featureless at all momenta. The extracted primary spectra along the band `B1' shown in the main text, as well as at three high-symmetry momentum values, continue to exhibit the peak-dip-hump structure. These results convincingly rule out the possibility that such spectral function anomaly  is extrinsic, but rather a manifestation of intrinsic many-body effects.

\section{Theory}

\subsection{Band structure calculations}
\begin{figure}[h]
\rotatebox[origin=c]{0}{\includegraphics[width=.95\columnwidth]{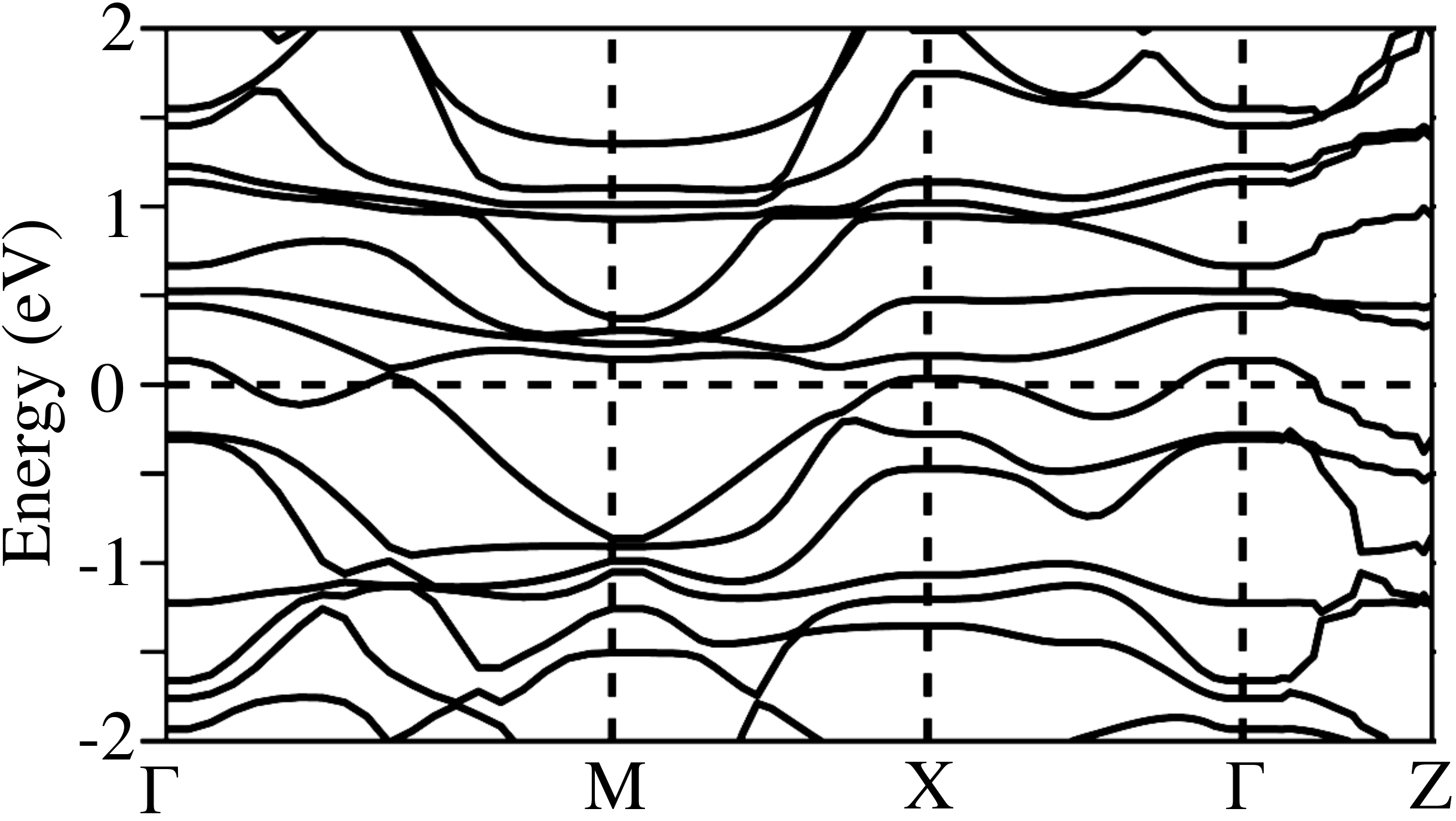}}
\caption{
 {\bf Fig. S4. GGA band structure.}
 Computed {\it ab-initio} band structure along high-symmetry line. The results agree well with previous calculation using full-potential code.\cite{Opahle2004}}
\end{figure}

We performed electronic structure calculations of UCoGa$_5$ within the framework of DFT and the results are shown in Fig.~S4. Our calculations were carried out by using the full-potential linearized augmented plane wave (FP-LAPW) method as implemented in the WIEN2k code~\cite{PBlaha2001}. The generalized gradient approximation (GGA)~\cite{JPPerdew1996} was used
for exchange-correlation functional. The spin-orbit coupling was included in a second variational way,
for which relativistic $p_{1/2}$ local orbitals were added into the basis set for the description of the $6p$ states of uranium \cite{Kunes2001}. The energy spread to separate the localized valence states was -6 Ryd. The muffin-tin  radii were: $2.5 a_0$ for U, $2.48 a_0$ for Co, and $2.2 a_0$ for Ga, where $a_0$ is the Bohr radius. The criterion for the number of plane waves was $R_{MT}^{{\rm min}} K^{{\rm max}} = 8$ and the number of $\mathbf{k}$-points was $40\times 40 \times 25$.

\subsection{Spin-fluctuation coupling calculation}
The non-interacting spin susceptibility for a multiband system is a tensor in the orbital basis \cite{Graser},  whose components are
\begin{widetext}
\begin{equation}\label{chibare}
\chi^0_{stpr}({\bm q},\Omega)=-\frac{1}{N}\sum_{{\bm k},n,m}\phi_{n}^s({\bm k})\phi_{n}^{t*}({\bm k})
\phi_{m}^p({\bm k}+{\bm q})\phi_{m}^{r*}({\bm k}+{\bm q})\frac{f\left(\xi_{n} ({\bm k})\right)-f\left(\xi_{m} ({\bm k}+{\bm q})\right)}{\Omega+i\delta-\left(\xi_{n} ({\bm k})-\xi_{m} ({\bm k}+{\bm q})\right)}.
\end{equation}
\end{widetext}
Here $\phi_{n}^i$ is the eigenstate for $n^{th}$ band $E_{n}$ projected on $i^{th}$-orbital, which are obtained directly from first-principles band-structure calculations. The susceptibility calculation is done with 40 bands that are present in the relevant $\pm$10~eV energy window around the Fermi level. The many-body corrections are obtained within the random-phase-approximation (RPA) given by
\begin{equation}
\hat{\chi}=\hat{\chi}^0[1-\hat{U} \hat{\chi}^0]^{-1}.
\end{equation}
The hat symbol $\hat{}$ over a quantitity means that it is a matrix of $N^2\times N^2$ size where $N$ is the number of orbitals.
The matrix $\hat{\chi}^0$ has elements $\chi^0_{stpr}$. The interaction matrix $\hat{U}$ is defined in the same basis consisting of intraorbital $U$, interorbital $V$, Hund's coupling $J$ and pair-scatterings $J^{\prime}$ terms \cite{Graser,Dastworesonance}.  In the present calculation, we will ignore for simplicity the matrix elements due to orbital overlap of the eigenstates, i.e., we assume $\phi^i_{n}=1$ when $i=n$. Such an approximation simplifies the calculation and $\hat{\chi}^0$ becomes a diagonal matrix with $J=J^{\prime}=V=0$, and $U=$1~eV.

In the fluctuation exchange approximation \cite{Bickers}, the matrix elements of the spin-fluctuation interaction vertex  $\hat{V}$ are given by
\begin{equation}
V_{stpr}({\bf q},\Omega) = \left[\frac{3}{2}\hat{U}\hat{\chi}^{\prime\prime}({\bm q},\Omega)\hat{U}\right]_{stpr}
\end{equation}
From this expression the self-energy is obtained by using the Feynman-Dyson formalism. When employing the spectral representation of the Matsubara Green's function matrix $\hat{G}$,
\begin{equation}
G_{st}({\bf k},i\omega_m) = \sum_{n}\frac{\phi_{n}^s({\bm k})\phi_{n}^{t*}({\bm k})}{i\omega_m-E_{n}({\bm k})}.
\end{equation}
We can write the self-energy within the GW approximation, using the fluctuation-exchange interaction matrix $\hat{V}$:
\begin{eqnarray}\label{FD}
{\rm Im} \, \Sigma_{st}({\bm k},\omega)
&=&\sum_{pr}\int\frac{d{\bm q}}{N}\int_{-\infty}^{\infty} d\Omega~[f_{st}({\bm k})+n_B(\Omega)] \Gamma({\bm k},{\bm q},\omega,\Omega)\nonumber\\
&&V^{stpr}_{eff}({\bm q},\Omega)~{\rm Im}~G_{pr}({\bm k}-{\bm q},\omega-\Omega) ,
\end{eqnarray}
The above equation denotes the quasiparticle excitation at the frequency $\omega$. To reduce the number of indices in the main text, we absorb the orbital indices $s$ and $t$ into its corresponding band index $n$:
$\Sigma_{st} \to \Sigma_{n}$.
 The electronic and bosonic occupation numbers are $f_{st}({\bm k})$ and $n_B(\Omega)$. The real part of the self-energy is obtained from Eq.~(\ref{FD}) by employing the Kramers-Kronig relationship. Combined with Dyson's equation
\begin{equation}
\hat{G}^{-1} = \hat{G}_0^{-1}-\hat{\Sigma} ,
\end{equation}
we obtain the many-body renormalized Green's function $\hat{G}$ matrix.
Full self-consistency requires the bare Green's function $\hat{G}_0$ to be replaced with self-energy dressed $\hat{G}$ in $\hat{\chi}^0$. This procedure is numerically expensive, especially in multiband systems. Therefore, we adopt a modified self-consistency scheme, where we expand the real part of the self-energy $\Sigma^{\prime}\approx(1-Z^{-1})\omega$ in the low-energy region, where $\Sigma^{\prime\prime}\approx0$. The resulting self-energy dressed quasiparticle dispersions ${\bar \xi}_{n}({\bm k})=Z\xi_{n}({\bm k})$ are used in Eq. (2)-(6), which keeps all the formalism unchanged with respect to $Z$. Finally, it introduces a vertex correction which simplifies to $\Gamma({\bm k},{\bm q},\omega,\Omega)=1/Z$ according to Ward's identity.

\section{Details of self-energy correction to individual bands}
%
%
\begin{figure}[h]
\rotatebox[origin=c]{0}{\includegraphics[width=.9\columnwidth]{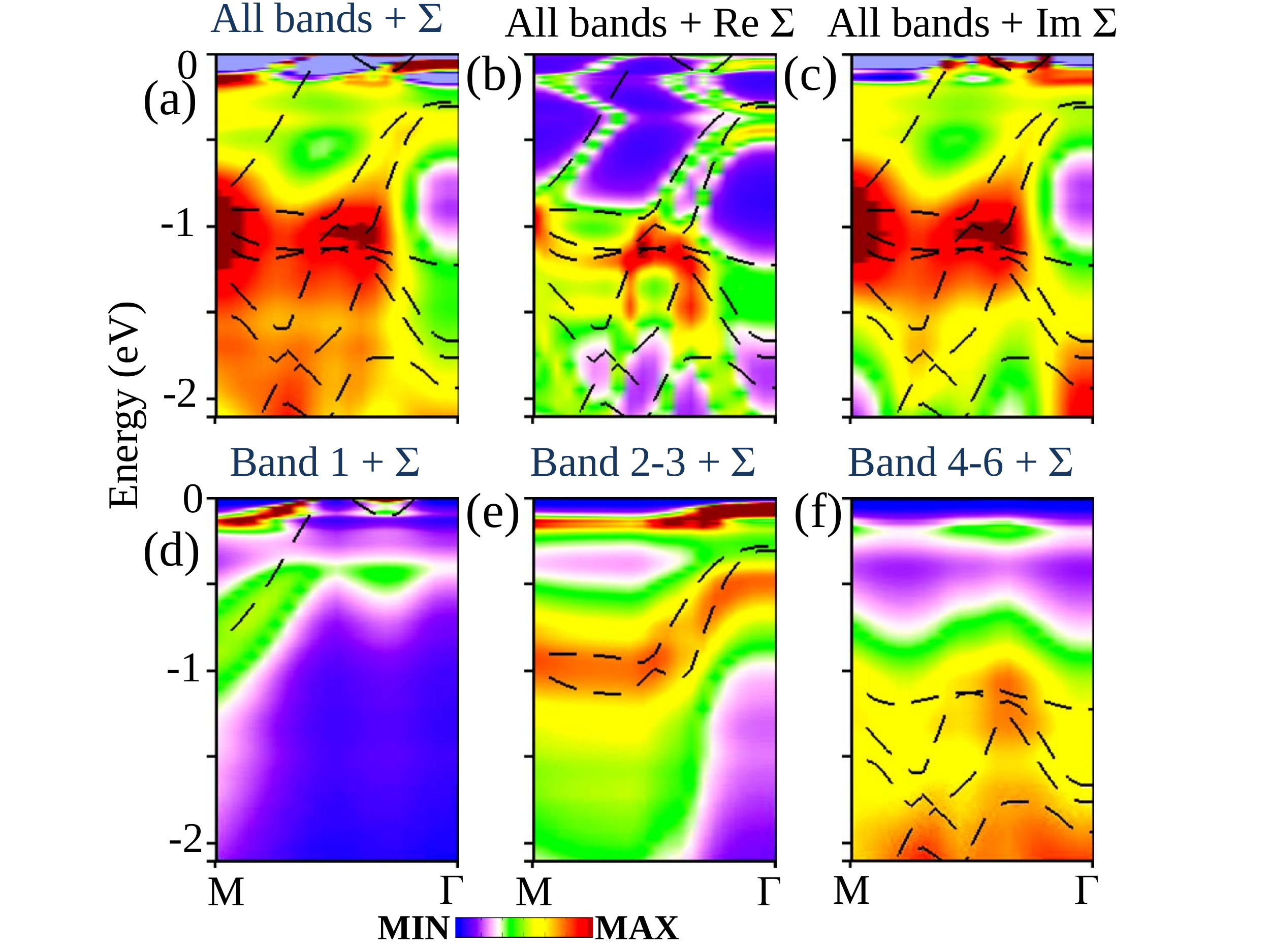}}
\caption{
{\bf Fig. S5. Computed self-energy corrections to individual bands.}
(a) The self-energy dressed spectral function along M$\rightarrow\Gamma$, replotted from Fig.~3(a) of the main text. The Re~$\Sigma$ [with Im~$\Sigma$ =constant] and Im~$\Sigma$  [Re~$\Sigma$=0] corrected spectral function plots are given in (b) and (c). The spectral function renormalization effect due to the full self-energy correction is shown individually for various bands in (d) through (f). Note that the self-energy itself is calculated for all bands and is kept same for all figure panels.}
\end{figure}

Figure~S5 demonstrates how different parts of the self-energy dress different electronic states. To make this comparative study easier, we use the same self-energy in all panels. As shown in Fig.~S5(b), the real part of the self-energy creates the dispersion renormalization which is strongly energy and band dependent. In this case we have used a constant broadening. The strong band renormalization in the low-energy region at M point is created by band `B1' [see Fig.~S5 by Re~$\Sigma$. By comparing Fig.~S5, (b) and (c), we notice that although there are renormalized bands present in the energy scale of $\sim$500~meV in 3(b), yet the imaginary part of the self-energy creates a dip at this energy as seen in S5(c). This result emphasizes the importance of Im~$\Sigma$ in that it  alone can produce the peak-dip-hump feature in the spectral function. This is also evident in Fig.~S5(e) and S5(f), where we see that despite the absence of low-energy quasiparticle bands, a nearly dispersionless band is created around -80 meV. For a purely momentum-independent  self-energy, the low-energy feature is completely dispersionless, while momentum dependence is introduced by the presence of bare electronic bands in this energy region, in addition to a momentum-dependent self-energy.

\section{Fermi surface topology renormalized by interaction and comparison with de-Haas van Alphen oscillations}
%
%
\begin{figure}[h]
\rotatebox[origin=c]{0}{\includegraphics[width=.99\columnwidth]{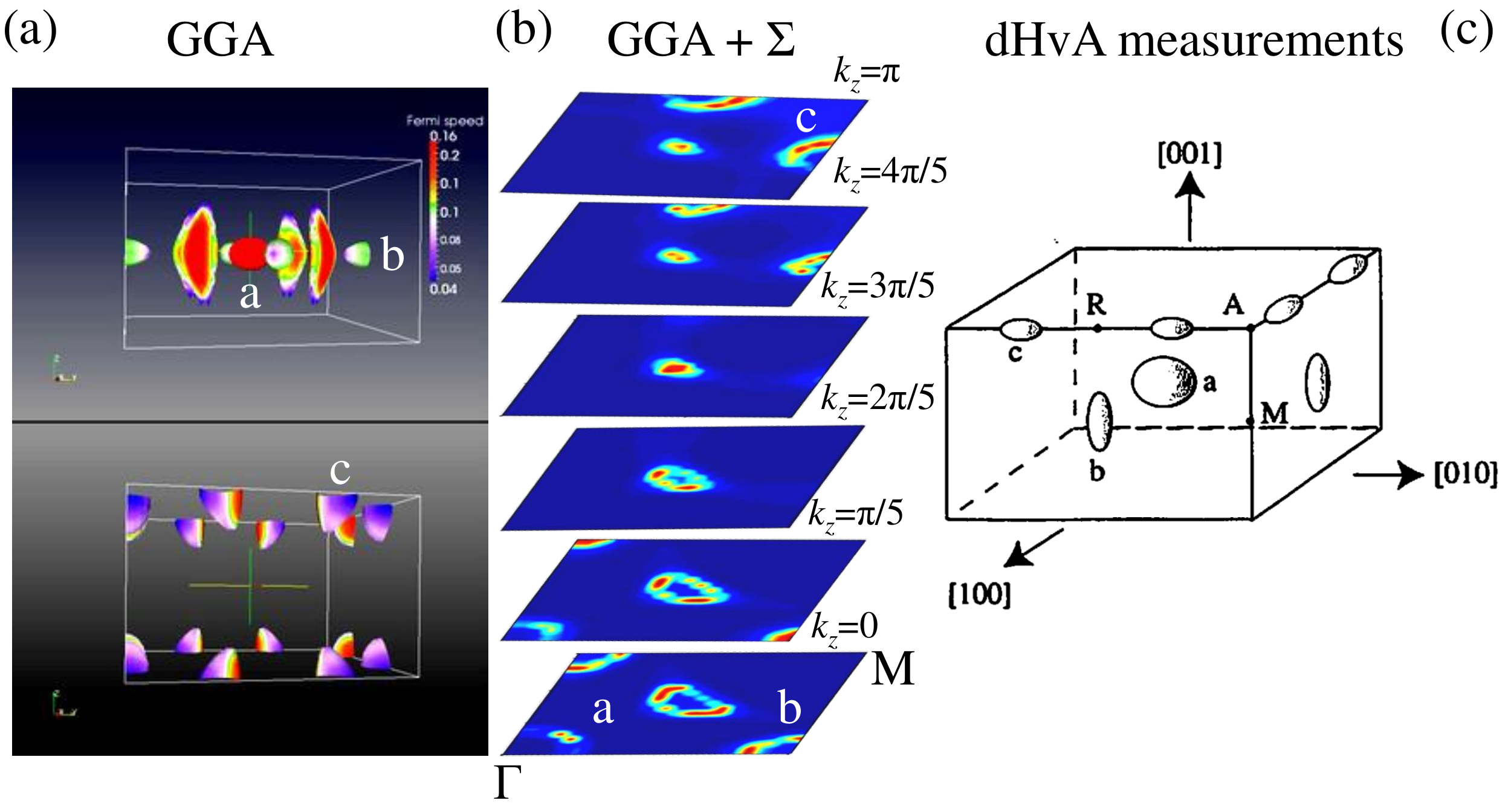}}
\caption{
{\bf Fig. S6. FS maps and dHvA results.} 
(a) The GGA calculated FSs for the two bands crossing the Fermi level. The color coding reflects the Fermi velocity or the inverse of the density of states. The box is centered at the $\Gamma$ point.
(b) FS cuts at six representative $k_z$-values after including the self-energy ($\Sigma$) correction. 
(c) Measured FS pockets from dHvA oscillations \cite{QO}. In the GGA,  GGA+$\Sigma$, and dHvA data, all the FS pieces labeled 
`a', `b', and `c' are well reproduced, while the large tube-like FSs `d' obtained in our calculation in (a) and (b) around  $(\pi/2,\pi/2,h)$  are not seen in the dHvA experiment in (c).
}
\end{figure}

In Fig.~S6, we present our theoretical Fermi surface (FS) maps calculated without self-energy correction in (a), with self-energy correction in (b), and compare with the measured FSs via the de-Haas van Alphen (dHvA) measurements by Ikeda {et al.} \cite{QO} in (c). In the $k_z=0$ plane, we obtain three FSs: at the $\Gamma$ point (labeled `a' FS), at $(\pi,0,0)$ (labeled `b' FS), and around $(\pi/2,\pi/2,h)$ connecting the $\Sigma$ to $S$ points  (labeled `d' FS), and their equivalent points in the Brillouin zone. In the $k_z=\pi$ plane, the `a' and `b' FSs disappeared, whereas the `c' FSs reach maximum size. The GGA results are in good agreement with previous electronic structure calculations by Opahle {\it et al.} \cite{Opahle2004}. After including the spin-fluctuation interaction, the spectral weight across each FS varies, but the general shape and position of each FS remains very much the same. Only the tube-shaped `d' pockets changed from closed to open orbits along the $k_z$ direction due to the small shift of the chemical potential to lower energies. This sensitivity to small changes in the electronic structure calculation is consistent with earlier reports of closed versus open tube-like pockets\cite{Maehira2003,Opahle2004}. The `a', `b' and `c' pockets, seen in the dHvA measurements, are in qualitative agreement with electronic structure calculations, as previously discussed  by Opahle {\it et al.} \cite{Opahle2004}, while the `d' FS was not observed. It is possible that the open-loop pocket `d' was masked by the relatively broad fundamental frequency of the `a' orbit. A targeted search for the associated dHvA oscillation should resolve the so far missing tube-shaped pocket.

\section{Concept of `waterfall' feature and its similarity with cuprates}
%
%
\begin{figure}[h]
\rotatebox[origin=c]{0}{\includegraphics[width=.9\columnwidth]{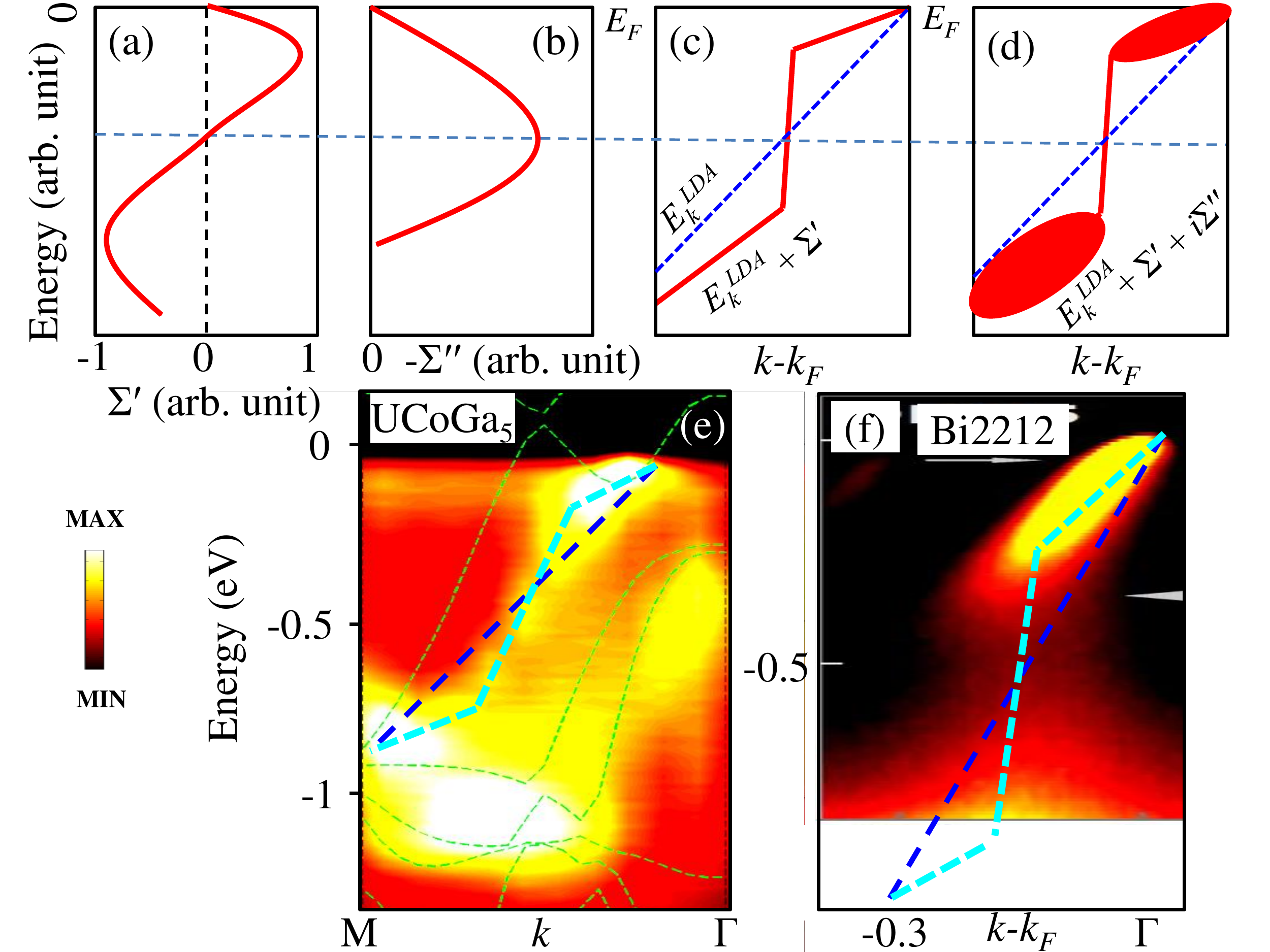}}
\caption{
{\bf Fig. S7. Illustration of band renormalization.} (a) Schematic energy dependence of Re~$\Sigma$, containing a sign-reversal at a characteristic energy. (b) Im~$\Sigma$ is expected to obtain a peak at the energy where Re~$\Sigma$ changes sign. (c) A linear dispersion is shown to be renormalized by Re~$\Sigma$. (d) Typical form of the spectral weight map in the energy and momentum space due to full $\Sigma$-correction. (e) ARPES data of UCoGa$_5$ presented in Fig.~1 of the main text (now replotted in different color map). (f) A typical `waterfall' data of Bi-based cuprate superconductor\cite{Graf2007} is shown here for comparison. The cyan dashed line guides the eyes on how the experimental dispersion is deviated from the linear bare dispersion (blue dashed line).}
\end{figure}

In Fig.~S7, we demonstrate the concept of `waterfall' physics. The basic concept that underlies this lineshape of quasiparticles is that the mass renormalization is strongly energy dependent  and there exists a crossover where the renormalization changes from positive to negative.  In this case, the strength of the Im~$\Sigma({\bm k},\omega)$ peak acquires values that split the overall quasiparticle states into two energy scales; at low energies it induces itinerant quasiparticle states, while at higher energies it creates incoherent localized states. These two energy scales yield a ubiquitous peak-dip-hump feature, which is consistent with photoemission data, see Fig. 1 for a detailed comparison. The resulting Re $\Sigma({\bm k},\omega)$ acquires an interesting energy dependence as shown in Fig. S7(a). The peak height of Im~$\Sigma({\bm k},\omega)$ is sufficiently large such that it leads to Re $\Sigma({\bm k},\omega)$ at this energy, which is possible within the intermediate Coulomb interaction scenario, but not the weak-coupling regime. Below this crossover energy Re $\Sigma({\bm k},\omega)>0$, which causes the quasiparticle to acquire a mass enhancement or reduction in quasiparticle weight, i.e.,  $Z<1$. As a consequence the corresponding bands are renormalized toward the Fermi level. Above this characteristic energy Re $\Sigma({\bm k},\omega)<0$, making the quasiparticle states shift toward higher energy ($Z>1)$. At the crossover energy, the dispersion renormalization is completely eliminated, making the overall quasiparticle state to sharply drop from $Z<1$ to $Z>1$ regions. At the crossover energy, the spectral weight is strongly suppressed due to the peak in  Im~$\Sigma({\bm k},\omega)$. The obtained shape of the quasiparticle dispersion is dubbed `waterfall' feature due to both its visual  and conceptual similarity.

To demonstrate that the ARPES data of UCoGa$_5$ indeed obtain similar `waterfall' feature as in cuprate superconductors, we replot our data in the same colorscale of the data available for Bi-based cuprate \cite{Graf2007}.  The characteristic similarity of the lineshape of the two spectra affirms our claim that a `waterfall' feature is present in actinide systems.

\end{document}